\documentclass[journal,comsoc]{IEEEtran}

\usepackage{graphicx}
\usepackage[T1]{fontenc}
\usepackage[bookmarks=false]{hyperref}

\usepackage{enumitem}
\usepackage[cmex10]{amsmath}
\usepackage{amssymb}
\usepackage{cite}
\usepackage{amsthm}
\usepackage{textcomp}
\usepackage{siunitx}
\usepackage{color}
\usepackage{subfigure}
\usepackage{cases}
\usepackage[font={small}]{caption}
\usepackage{bbm}
\usepackage{tcolorbox}

\theoremstyle{definition}

\theoremstyle{remark}

\setlist[description]{style=multiline}

\IEEEoverridecommandlockouts

\begin{document}
\sloppy

\title{Coding for Distributed Fog Computing} 

\author{Songze~Li,
        Mohammad~Ali~Maddah-Ali,
        and~A.~Salman~Avestimehr
\thanks{S.~Li and A.S.~Avestimehr are with the Department of Electrical Engineering, University of Southern California, Los Angeles, CA, 90089, USA (e-mail: songzeli@usc.edu; avestimehr@ee.usc.edu).}
\thanks{M. A. Maddah-Ali is with Nokia Bell Labs, Holmdel, NJ, 07733, USA (e-mail: mohammad.maddahali@nokia-bell-labs.com).}
}

\maketitle

\section{abstract}
Redundancy is abundant in Fog networks (i.e., many computing and storage points) and grows linearly with network size. We demonstrate the transformational role of coding in Fog computing for leveraging such redundancy to substantially reduce the bandwidth consumption and latency of computing. In particular, we discuss two recently proposed coding concepts, namely Minimum Bandwidth Codes and Minimum Latency Codes, and illustrate their impacts in Fog computing. We also review a unified coding framework that includes the above two coding techniques as special cases, and enables a tradeoff between computation latency and communication load to optimize system performance. At the end, we will discuss several open problems and future research directions.

\section{Introduction}\label{sec:intro}
\begin{figure}[htbp]
   \centering
   \includegraphics[width=0.45\textwidth]{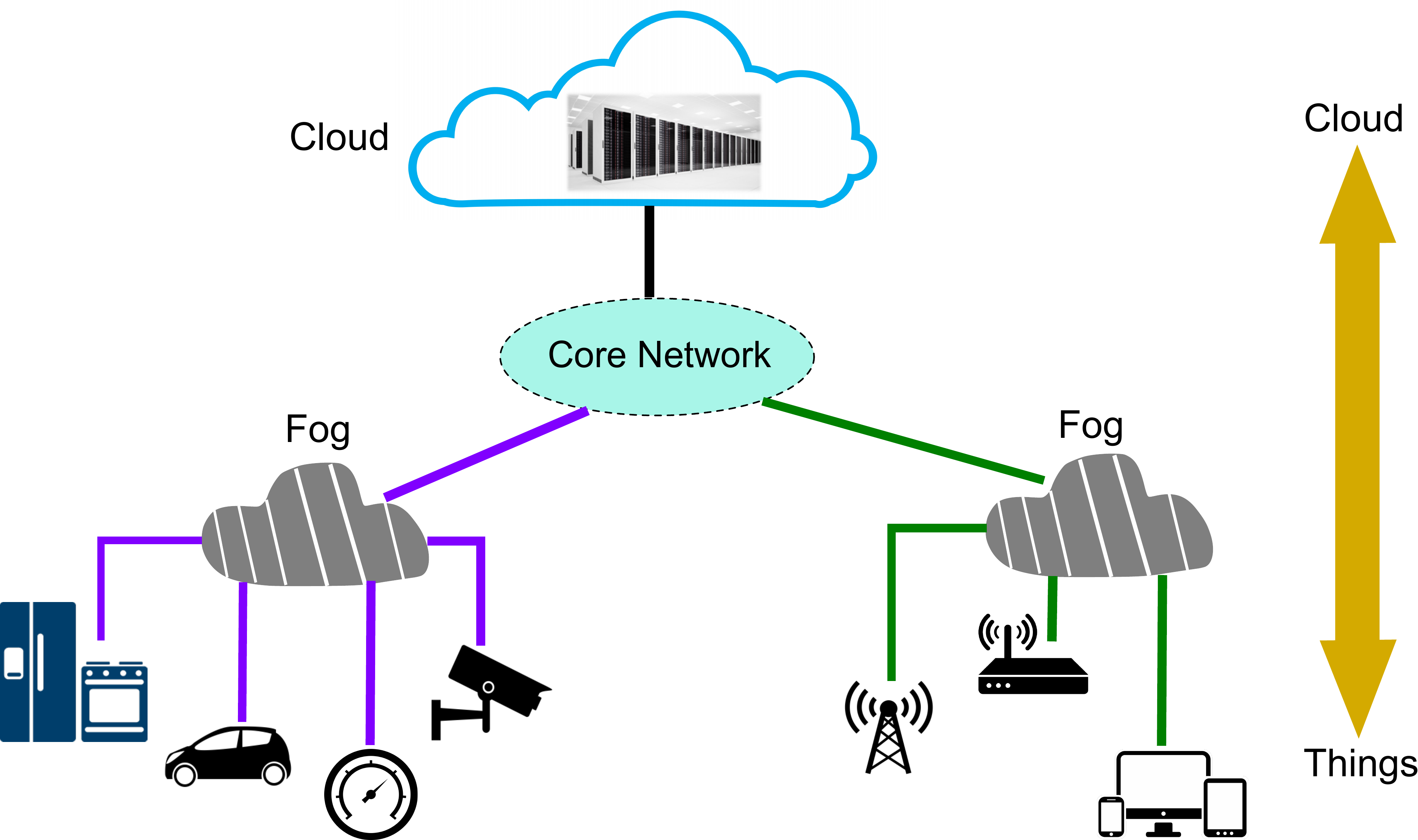}
   \caption{Illustration of a Fog architecture.}
   \label{fig:Fog}
\end{figure}

The Fog architecture (see Fig.~\ref{fig:Fog}) has been recently proposed to better satisfy the service requirements of the emerging Internet-of-Things (IoT) (see, e.g.,~\cite{bonomi2012fog}). Unlike the Cloud computing that stores and processes end-users' data in remote and centralized datacenters, Fog computing brings the provision of services closer to the end-users by pooling the available resources at the edge of the network (e.g., smartphones, tablets, smart cars, base stations and routers) (see, e.g.,~\cite{CZ,yi2015survey}). As a result, the main driving vision for Fog computing is to leverage the significant amount of dispersed computing resources at the edge of the network to provide much more user-aware, resource-efficient, scalable and low-latency services for IoT.

The main goal of this paper is to demonstrate how coding can be effectively utilized to trade abundant computing resources at network edge for communication bandwidth and latency. In particular, we illustrate two recently proposed novel coding concepts that leverage the available or under-utilized computing resources at various parts of the network to enable coding opportunities that significantly reduce the bandwidth consumption and latency of computing, which are of particular importance in Fog computing applications.

The first coding concept introduced in~\cite{LMA_all,li2016fundamental}, which we refer to as \emph{Minimum Bandwidth Codes}, enables a surprising inverse-linear tradeoff between computation load and communication load in distributed computing. Minimum Bandwidth Codes  demonstrate that increasing the computation load by a factor of $r$ (i.e., evaluating each computation at $r$ carefully chosen nodes) can create novel coding opportunities  that reduce the required communication load for computing by the same factor. Hence, Minimum Bandwidth Codes can be utilized to pool the underutilized computing resources at network edge to slash the communication load of Fog computing.

The second coding concept introduced in~\cite{lee2015speeding}, which we refer to as \emph{Minimum Latency Codes}, enables an inverse-linear tradeoff between computation load  and computation latency (i.e., the overall job response time). More specifically, Minimum Latency Codes utilize coding to effectively inject redundant computations to alleviate the effects of stragglers and speed up the computations by a multiplicative factor that is proportional to the amount of injected redundancy. Hence, by utilizing more computation resources at network edge, Minimum Latency Codes can significantly speed up distributed Fog computing applications.   

In this paper, we give an overview of these two coding concepts, illustrate their key ideas via motivating examples, and demonstrate their impacts on Fog networks. 
More generally, noting that redundancy is abundant in Fog networks  (i.e., many computing/storage points) and grows linearly with network size, we demonstrate the transformational role of coding in Fog computing for leveraging such redundancy to substantially reduce the bandwidth consumption and latency of computing. We also point out that while these two coding techniques are also applicable to Cloud computing applications, they are expected to play a much more substantial role in improving the system performance of Fog applications, due to the fact that communication bottleneck and straggling nodes are far more severe issues in Fog computing compared with its Cloud counterpart.

We also discuss a recently proposed \emph{unified} coding framework, in which the above two coding concepts are systematically combined by introducing a tradeoff between ``computation latency'' and ``communication load''. This framework allows a Fog computing system to operate at any point on the tradeoff, on which the Minimum Bandwidth Codes  and the Minimum Latency Codes  can be viewed as two extreme points that respectively minimizes the communication load and the computation latency. 

We finally conclude the paper and highlight some exciting open problems and research directions for utilizing coding in Fog computing architectures.

\section{Minimum Bandwidth Codes}
We illustrate Minimum Bandwidth Codes in a typical Fog computing scenario, in which a Fog client aims to utilize the network edge for its computation task.  For instance, a driver wants to find the best route through a navigation application offered by the Fog, in which the map information and traffic condition are distributedly stored in edge nodes (ENs) like roadside monitors, smart traffic lights, or other smart cars that collaborate to find the best route. Another example is object recognition that is the key enabler of many augmented reality applications. To provide an object recognition service over Fog, edge nodes like routers and base stations, each stores parts of the dataset repository, and collaboratively process the images or videos provided by the Fog client. 

For the above Fog computing applications, the computation task is over a large dataset that is distributedly stored on the edge nodes (e.g., map/traffic information or dataset repository), and  the computations are often decomposed using MapReduce-type frameworks (e.g.,~\cite{dean2004mapreduce,zaharia2010spark}), in which a collection of edge nodes distributedly \emph{Map} a set of input files, generating some intermediate values, from which they \emph{Reduce} a set of output functions. 

\begin{figure}[t]
   \centering
\subfigure[]{\includegraphics[width=0.48\textwidth]{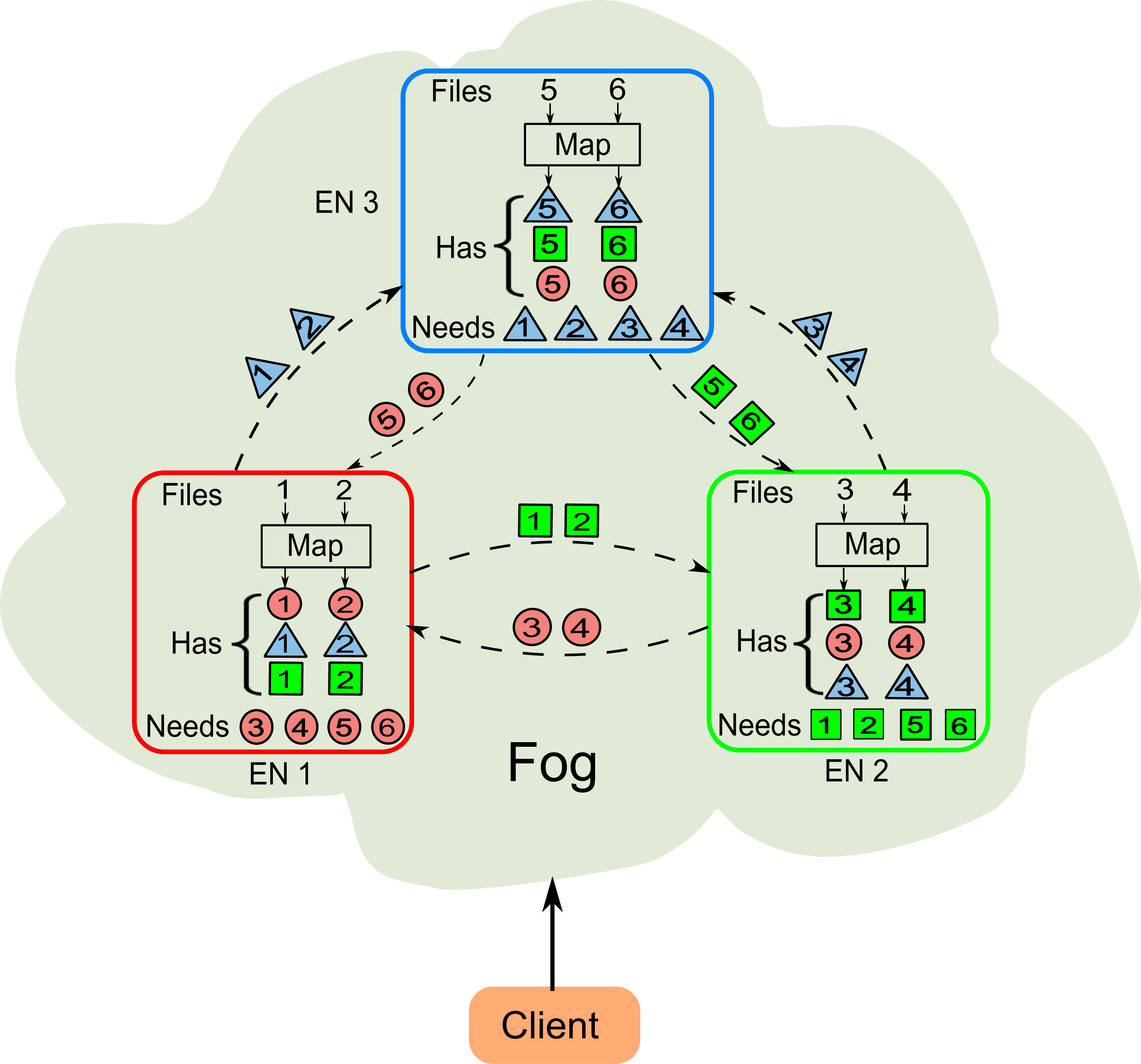}
       \label{fig:MBC-uncoded}}
 \subfigure[]{\includegraphics[width=0.48\textwidth]{MBC-coded.pdf}
       \label{fig:MBC-coded}}
   \caption{a) An uncoded Fog computing scheme to compute 3 functions, one on each of the 3 ENs, from 6 files. Each file is mapped once on one EN, and each EN has 4 intermediate values transferred uncodedly from the other ENs to reduce the corresponding output; b) Implementation of a Minimum Bandwidth Code on 3 ENs. Each of the $6$ files is mapped on two ENs. During data shuffling, each EN creates a coded packet that is simultaneously useful for the other two ENs, by XORing two locally computed intermediate values, and multicasts the packet to the other two ENs.}
   \label{fig:MBC}
\end{figure}

We now  demonstrate the main concepts of Minimum Bandwidth Codes in a simple problem depicted in Fig.~\ref{fig:MBC}. In this case, a client uploads a job of computing $3$ output functions (represented by red/circle, green/square, and blue/triangle respectively) from $6$ input files to the Fog. Three edge nodes in the Fog, i.e., EN~$1$, EN~$2$ and EN~$3$, collaborate to perform the computation. Each EN is responsible for computing a unique output function, e.g., EN~$1$ computes the red/circle function, EN~$2$ computes the green/square function, and EN~$3$ computes the blue/triangle function. When an EN maps a locally stored input file, it computes $3$ intermediate values, one for each output function. To reduce an output function, each EN needs to know the intermediate values of this output for all $6$ input files.    

We first consider the case where no redundancy is imposed on the computations, i.e., each file is mapped exactly once. Then as shown in Fig.~\ref{fig:MBC-uncoded}, each EN maps $2$ input files locally, obtaining $2$ out of $6$ required intermediate values. Hence, each EN needs another $4$ intermediate values transferred from the other ENs, yielding a communication load of $4 \times 3=12$. 

Now, we demonstrate how Minimum Bandwidth Codes can substantially reduce the communication load by injecting redundancy in computation. As shown in Fig.~\ref{fig:MBC-coded}, let us double the computation such that each file is mapped on two ENs (files are downloaded to the ENs offline). It is apparent that since more local computations are performed, each EN now only requires $2$ other intermediate values, and an uncoded shuffling scheme would achieve a communication load of $2 \times 3=6$. However, we can do better with the Minimum Bandwidth Codes. As shown in Fig.~\ref{fig:MBC-coded}, instead of unicasting individual intermediate values, every EN multicasts a bit-wise XOR, denoted by $\oplus$, of $2$ intermediate values to the other two ENs, simultaneously satisfying their data demands. For example, knowing the blue triangle in File~$3$, EN~$2$ can cancel it from the coded packet multicast by EN~$1$, recovering the needed green square in File~$1$. In general, the bandwidth consumption of multicasting one packet to two nodes is less than that of unicasting two packets, and here we consider a scenario in which it is as much as that of unicasting one packet (which is the case for wireless networks).  Therefore, the above Minimum Bandwidth Code incurs a communication load of $3$, achieving a $4\times$ gain from the case without computation redundancy and a $2\times$ gain from the uncoded shuffling. 

More generally, we can consider a Fog computing scenario, in which $K$ edge nodes collaborate to compute $Q$ output functions from $N$ input files that are distributedly stored at the nodes. We define the computation load, $r$, to be the total number of input files that are mapped across the nodes, normalized by $N$. That is, e.g., $r = 2$ means that on average each file is mapped on two nodes. We can similarly define the communication load $L$ to be the total (normalized) number of information bits exchanged across nodes during data shuffling, in order to compute the $Q$ output functions.

\begin{figure}[htbp]
   \centering
   \includegraphics[width=0.4\textwidth]{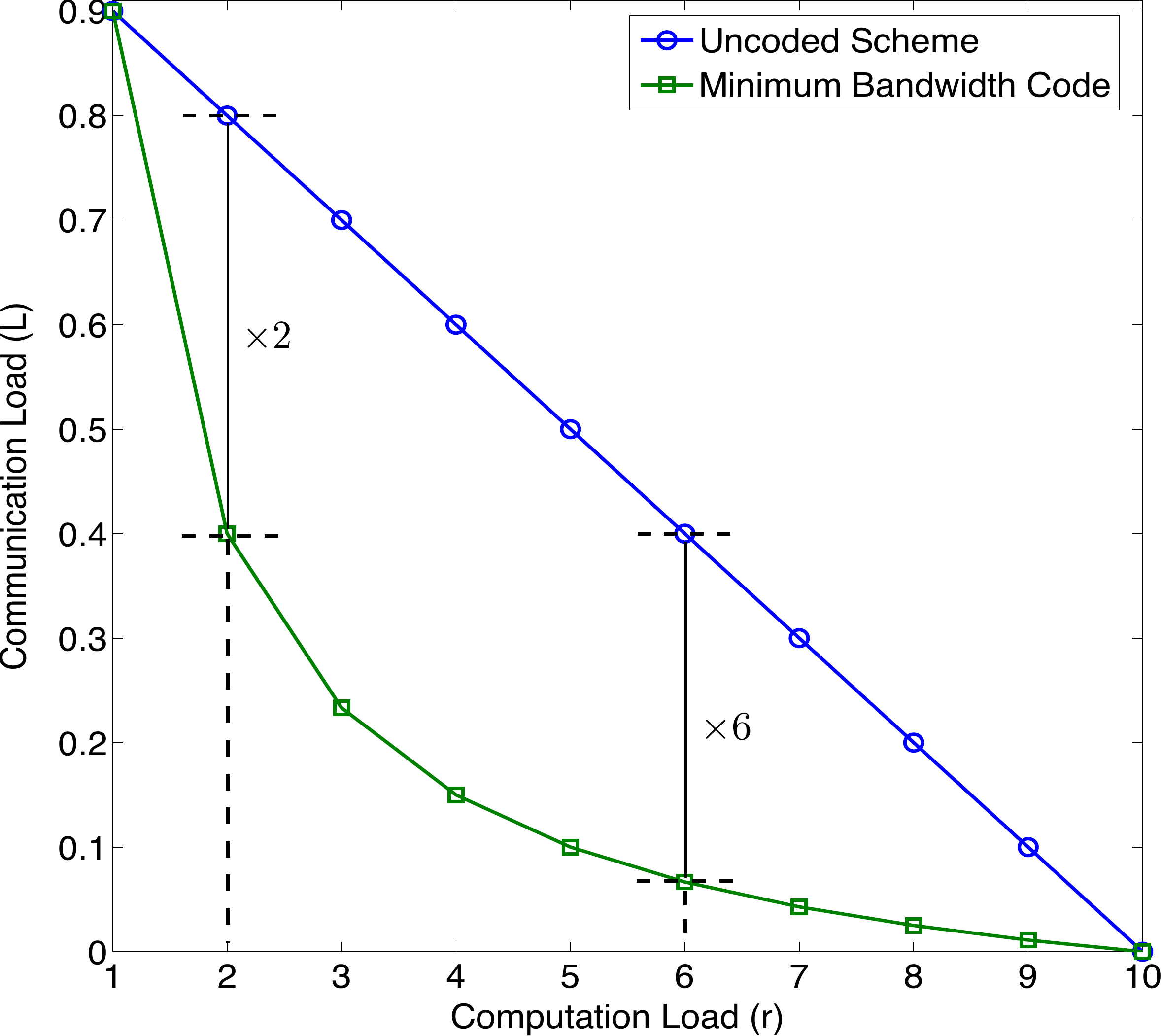}
   \caption{Comparison of the communication load of Minimum Bandwidth Codes with that of the uncoded scheme, for a network with $K=10$ edge nodes.}
   \label{fig:scaling}
\end{figure}

For this scenario it was shown in~\cite{li2016fundamental} that, compared with conventional uncoded strategies, Minimum Bandwidth Codes can surprisingly reduce the communication load by a multiplicative factor that equals to the computation load $r$, when computing $r$ times more sub-tasks than the execution without redundancy (i.e., $r=1$). Or more specifically,
\begin{align}
L_{\textup{coded}}= \tfrac{1}{r}L_{\textup{uncoded}}=\tfrac{1}{r}(1-\tfrac{r}{K})= \Theta(\tfrac{1}{r}).\label{eq:tradeoff}
\end{align}
Minimum Bandwidth Codes employ a specific strategy to assign the computations of the Map and Reduce functions, in order to enable novel coding opportunities for data shuffling. In particular, each data block is repetitively mapped on $r$ distinct nodes according to a specific pattern, in order to create coded multicast messages that deliver useful data simultaneously to $r \geq 1$ nodes. For example, as demonstrated in Fig.~\ref{fig:scaling}, the overall communication load can be reduced by more than 50\% when each Map task is repeated at only one other node (i.e., $r = 2$). 

The idea of efficiently creating and exploiting coded multicast opportunities was initially proposed to solve caching problems in~\cite{maddah2014fundamental,maddah2013decentralized}, and extended to wireless D2D networks in~\cite{ji2014fundamental}, where caches pre-fetch part of the content to enable coding during the content delivery, minimizing the network traffic. Minimum Bandwidth Codes extend such coding opportunities to data shuffling of distributed computing frameworks, significantly reducing the required communication load.

Apart from significantly slashing the bandwidth consumption, Minimum Bandwidth Codes also have the following major impacts on the design of Fog computing architecture.

{\bf Reducing Overall Response Time.} Let us consider an arbitrary Fog computing application for which the overall response time is composed of the time spent computing the intermediate tasks, denoted by $T_{\textup{Task Computation}}$, and the time spent moving intermediate results, denoted by $T_{\textup{Data Movement}}$. In many applications of interest (e.g., video/image analytics or recommendation services), most of the job execution time is spent for data movement. For example, consider the scenarios in which $T_{\textup{Data Movement}}$ is $10\times \sim 100\times$ of $T_{\textup{Task Computation}}$. Using a Minimum Bandwidth Code with computation load $r$, we can achieve an overall response time of 
\begin{align}
T_{\textup{total, coded}} \approx \mathbb{E}[rT_{\textup{Task Computation}}+ \tfrac{1}{r}T_{\textup{Data Movement}}].
\end{align}
To minimize the above response time, one would choose the optimum computation load $r^* = \sqrt{\frac{T_{\textup{Data Movement}}}{T_{\textup{Task Computation}}}}$.Then in the above example, utilizing Minimum Bandwidth Codes can reduce the overall job response time by approximately $1.5\sim5$ times. 

The impact of Minimum Bandwidth Codes on reducing the response time has been recently demonstrated in ~\cite{CTS16} through a series of experiments over Amazon EC2 clusters. In particular, the Minimum Bandwidth Codes were incorporated into the well-known distributed sorting algorithm \texttt{TeraSort}~\cite{o2008terabyte}, to develop a new coded sorting algorithm, namely \texttt{CodedTeraSort}, which allows a flexible selection of the computation load $r$. Here we summarize in Table~\ref{table:compare}, the runtime performance of a particular job of sorting 12 GB of data over 16 EC2 instances. 

\begin{table*}[!t]
\centering
\caption{Average response times for sorting 12 GB of data over 16 EC2 instances using 100 Mbps network speed.}
  \label{table:compare}
  \begin{tabular}{|c|c|c|c|c|c|c|c|c|}
    \hline
& CodeGen & Map & Pack/Encode & Shuffle & Unpack/Decode & Reduce & Total Time & Speedup \\
& (sec.)   & (sec.) & (sec.) & (sec.) & (sec.) & (sec.) & (sec.) & \\ \hline
    \texttt{TeraSort}:  &  --   & 1.86  & 2.35 & 945.72  & 0.85 & 10.47 & 961.25 & \\
    \texttt{CodedTeraSort}: $r=5$& 23.47 & 10.84 & 8.10 & 222.83  & 3.69 & 14.40 & 283.33 & 3.39$\times$ \\\hline
  \end{tabular}
\end{table*}

Theoretically according to~(\ref{eq:tradeoff}), with a computation load $r=5$, \texttt{CodedTeraSort} promises to reduce the data shuffling time by a factor of approximately $5$. From Table~\ref{table:compare}, we can see that while computing $r=5$ times more Map functions increased the Map task computation time by $5.83\times$, \texttt{CodedTeraSort} brought down the data shuffling time, which was the limiting component of the runtime of this application, by $4.24 \times$. As a result, \texttt{CodedTeraSort} reduced the overall job response time by $3.39\times$. 

{\bf Scalable Mobile Computation.} The Minimum Bandwidth Codes also found their application in a wireless distributed computing platform proposed in~\cite{li2016scalable}, which is a fully decentralized Fog computing environment. In this platform, a collection of mobile users, each has a input to process overall a large dataset (e.g., the image repository of an image recognition application), collaborate to store the dataset and perform the computations, using their own storage and computing resources. All participating users communicate the locally computed intermediate results among each other to reduce the final outputs. 

Utilizing Minimum Bandwidth Codes in this wireless computing platform leads to a \emph{scalable} design. More specifically, let us consider a scenario where $K$ users, each processing a fraction of the dataset, denoted by $\mu$ (for some $\frac{1}{K} \leq \mu \leq 1$), collaborate for wireless distributed computing. It is demonstrated in~\cite{li2016scalable} that Minimum Bandwidth Codes can achieve a (normalized) bandwidth consumption of $\frac{1}{\mu}-1$ to shuffle all required intermediate results. This reduces the communication load of the uncoded scheme, i.e. $K(1-\mu)$,  by a factor of $\mu K$, which scales linearly with the  aggregated storage size of all collaborating users. Also, since the consumed bandwidth is independent of the number of users $K$, Minimum Bandwidth Code allows this platform to simultaneously serve an unlimited number of users with a constant communication load.

\section{Minimum Latency Codes}
We now move to the second coding concept, named Minimum Latency Codes, and demonstrate it for a class of Fog computing applications, in which a client's input is processed over a large dataset (possibly over multiple iterations). The application is supported by a group of edge nodes, which have distributedly stored the entire dataset. Each node processes the client's input using the parts of the dataset it locally has, and returns the computed results to the client. The client reduces the final results after collecting intermediate results from all edge nodes. Many distributed machine learning algorithms fall into this category. For example, a gradient decent algorithm for linear regression requires multiplying the weight vector with the data matrix in each iteration. To do that at network edge, each edge node stores locally a sub-matrix of the data matrix. During computation, each edge node multiplies the weight vector with the stored sub-matrix and returns the results to the client.   

\begin{figure}[htbp]
   \centering
   \includegraphics[width=0.48\textwidth]{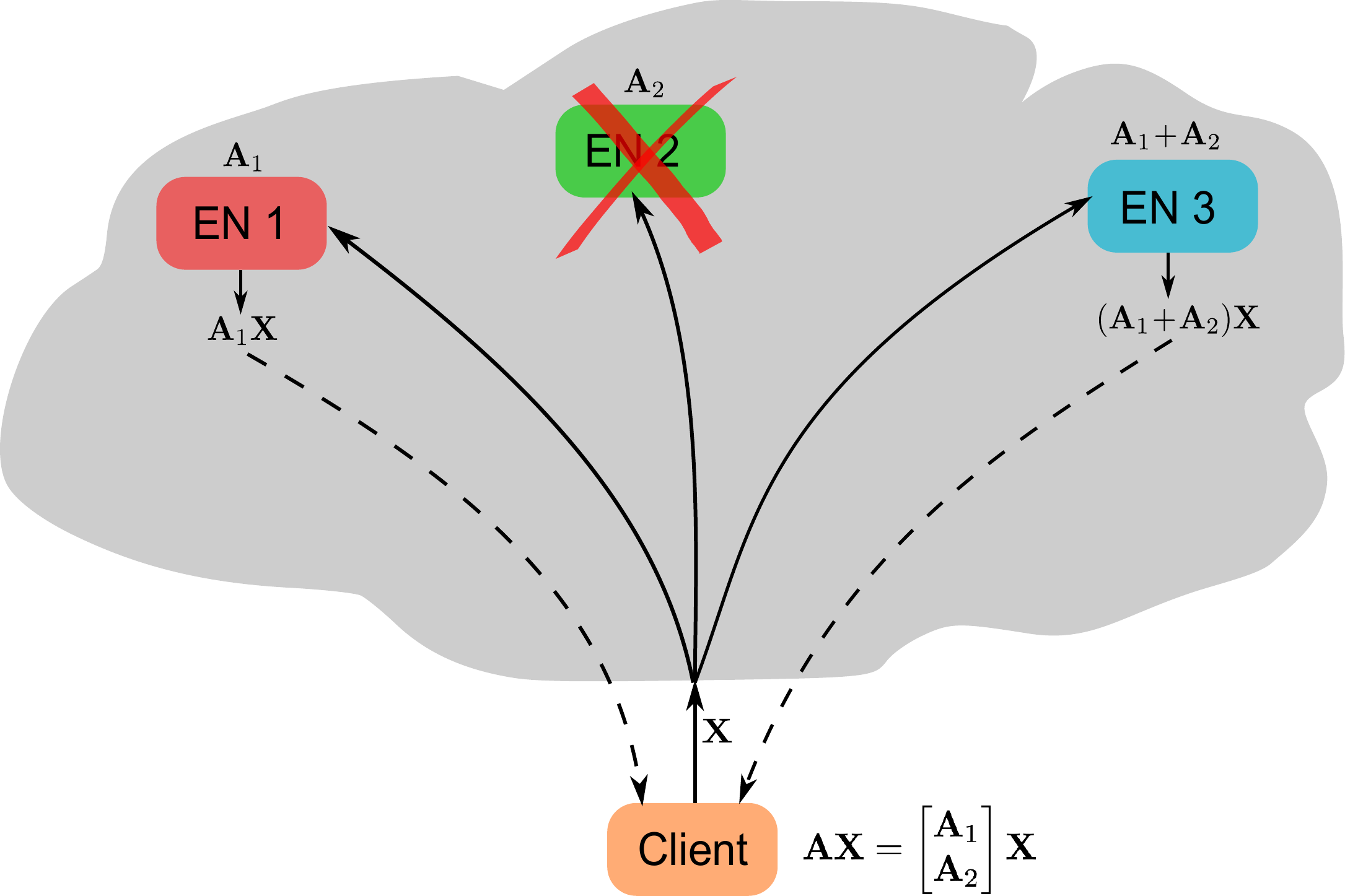}
   \caption{Matrix multiplication using a Minimum Latency Code. The code generates 3 coded tasks each executed by an EN. The client recovers the final result after receiving the results from EN $1$ and $3$. The runtime is not affected by the straggling EN~$2$.}
   \label{fig:MLC-coded}
\end{figure}

To be more specific, let us consider a simple distributed matrix multiplication problem, in which as shown in Fig.~\ref{fig:MLC-coded}, a client wants to multiply a data matrix ${\bf A}$ with the input matrix ${\bf X}$ to compute ${\bf A}{\bf X}$. The data matrix ${\bf A}$ is stored distributedly across $3$ nearby ENs, i.e., EN~$1$, EN~$2$, and EN~$3$, on which the matrix multiplication will be executed distributedly. 

One natural approach to tackle this problem is to vertically and evenly divide the data matrix ${\bf A}$ 
into $3$ sub-matrices, each of which is stored on one EN. Then when each EN receives the input ${\bf X}$, it simply multiplies its locally stored sub-matrix with ${\bf X}$ and returns the results, and the client vertically concatenates the returned matrices to obtain the final result. However, we note that since this uncoded approach relies on successfully retrieving the task results from all $3$ ENs, it has a major drawback that once one of the ENs runs slow or gets disconnected, the computation may take very long or even fail to finish. Minimum Latency Codes deal with slow or unreliable edge nodes by optimally creating redundant computations tasks. As shown in Fig.~\ref{fig:MLC-coded}, a Minimum Latency Code vertically partitions the data matrix ${\bf A}$ into $2$ sub-matrices ${\bf A}_1$ and ${\bf A}_2$, and creates one redundant task by summing ${\bf A}_1$ and ${\bf A}_2$. Then ${\bf A}_1$, ${\bf A}_2$ and ${\bf A}_1+{\bf A}_2$ are stored on EN~$1$, EN~$2$, and EN~$3$ respectively. In the case of Fig.~\ref{fig:MLC-coded}, the computation is completed when the client has received the task results only from EN~$1$ and $3$, from which ${\bf A}_2{\bf X}$ can be decoded. In fact, it is obvious that the client can recover the final result once she receives the task results from any 2 out of the 3 ENs, without needing to wait for the slow/unreachable EN (EN~$2$ in this case). In summary, Minimum Latency Codes create redundant computation tasks across Fog networks, such that having \emph{any} set of certain number of task results is sufficient to accomplish the overall computation. Hence, applying Minimum Latency Codes on the abundant edge nodes can effectively alleviate the effect of stragglers and significantly speed up Fog computing. 

As illustrated in the above example, the basic idea of Minimum Latency Codes is to apply erasure codes on computation tasks, creating redundant coded tasks that provide robustness to straggling edge nodes. Erasure codes have been widely exploited to combat symbol losses in communication systems and disk failures in distributed storage systems. The simplest form of erasure codes, i.e., the repetition code, repeats each information symbol multiple times, such that a information symbol can be successfully recovered as long as at least one of the repeats survives. For example, modern distributed files systems like Hadoop Distributed File System (HDFS) replicates each data block three times across different storage nodes. Another type of erasure code, known as the Maximum-Distance-Separable (MDS) code, provides better robustness to erasures. An $(n,k)$ MDS code takes $k$ information symbols and encodes them into $n \geq k$ coded symbols, such that obtaining \emph{any} $k$ out of the $n$ coded symbols is sufficient to decode all $k$ information symbols. This ``any $k$ of $n$'' property is highly desirable due to the randomness of erasures. A successful application of the MDS code is the Reed-Solomon Code used to protect CDs and DVDs.

As introduced in~\cite{lee2015speeding}, Minimum Latency Codes are exactly MDS codes that are used to encode computation tasks. For a Fog computing job executed on $n$ edge nodes, a $(n,k)$ Minimum Latency Code first decomposes the overall computation into $k$ smaller tasks, for some $k \leq n$. Then it encodes them into $n$ coded tasks using an $(n,k)$ MDS code, and assigns each of them to a node to compute. By the aforementioned ``any $k$ of $n$'' property of the MDS code, we can accomplish the overall computation once we have collected the results from the \emph{fastest} $k$ out of $n$ coded tasks, without worrying the tasks still running on the slow nodes (or stragglers). 

Minimum Latency Codes can help to significantly improve the response time of Fog applications. Let's consider a computation task performed distributedly across $n$ edge nodes. The response time of the uncoded approach is limited by the slowest node. An $(n,k)$ repetition code breaks the computation into $k$ tasks, and repeats each task $\frac{n}{k}$ times across the $n$ nodes, and the computation continues until each task has been computed at least once. On the other hand, for an $(n,k)$ Minimum Latency Code, the response time is limited by the fastest $k$ out of $n$ nodes that have finished their coded tasks. As shown in~\cite{lee2015speeding}, for a shifted-exponential distribution, the average response times of the uncoded execution and the repetition code are both $\Theta(\frac{\log n}{n})$. The Minimum Latency Codes can reduce the response time by a factor of $\Theta(\log n)$. For example, in a typical Fog computing scenario with $10 \sim 100$ nodes, Minimum Latency Codes can theoretically offer a $2.3\times \sim 4.6 \times$ speedup. Moreover, experiments on Amazon EC2 clusters were performed in~\cite{lee2015speeding}, in which for a gradient descent computation for linear regression, Minimum Latency Codes reduce the response time by 35.7\% on average. We further envision that in a Fog computing environment where computing nodes are much more heterogeneous and likely to be irresponsive, the performance gain by using Minimum Latency Codes will be much larger.

Other than speeding up the Fog computing applications, Minimum Latency Codes also maximize the survivability of the computation when faced with nodes failure/disconnection, i.e., when the task results may never come back. We note that an $(n,k)$  Minimum Latency Code requires any $k$ out of $n$ tasks to be returned to guarantee a successful computation, and this level of robustness can not be provided by either the uncoded computation or the the repetition code. 

\section{A Unified Coding Framework}
We have so far discussed two different coding techniques that aim at minimizing the bandwidth consumption and the computation latency of Fog computing. However, under a MapReduce-type computing model, a \emph{unified} coded framework has been recently developed in~\cite{LMA16_unify} by introducing a tradeoff between ``computation latency'' in the Map phase and ``communication load'' in the Shuffle phase. 

\begin{figure}[htbp]
   \centering
   \includegraphics[width=0.4\textwidth]{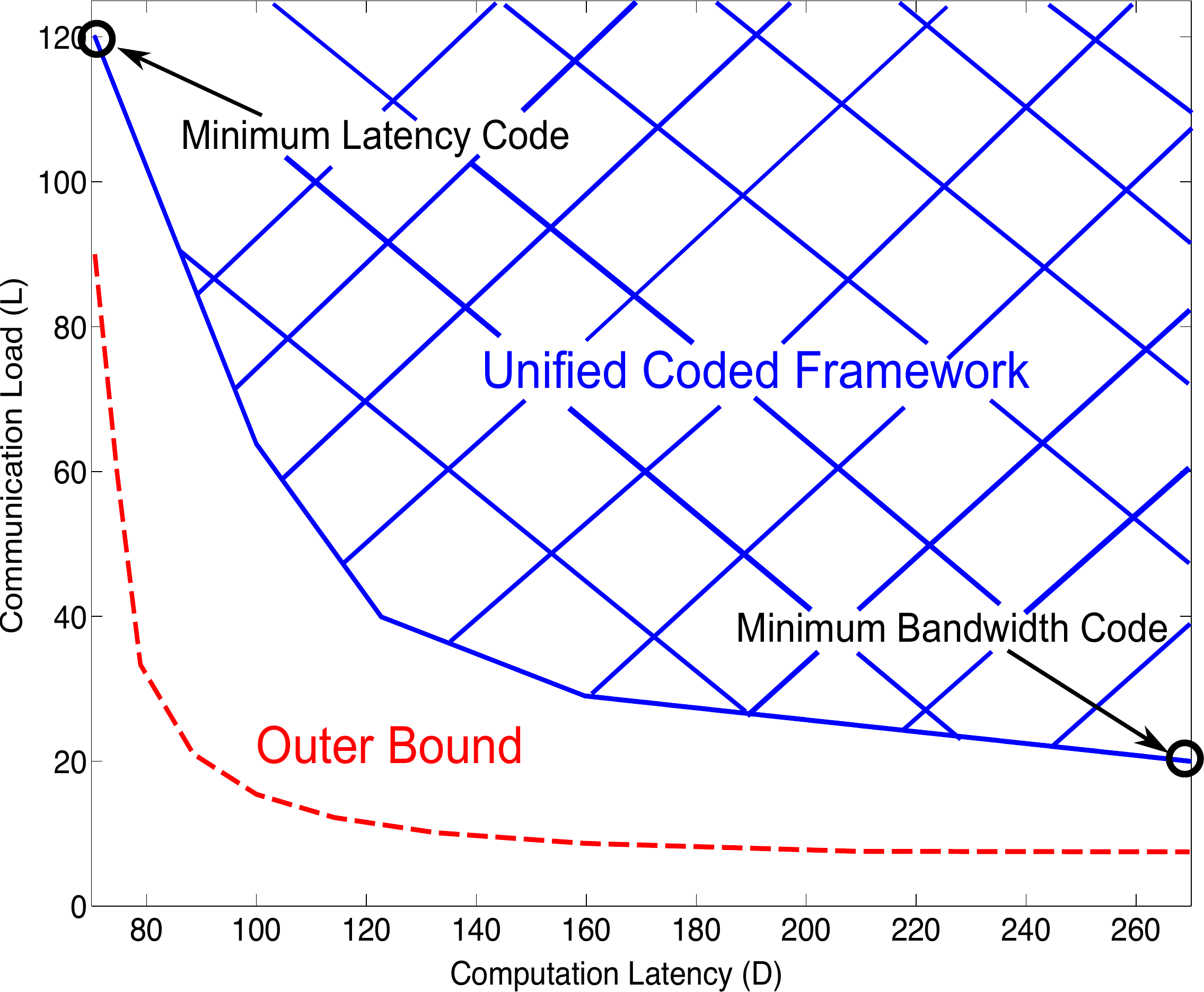}
   \caption{Comparison of the latency-load pairs achieved by the proposed unified scheme with the outer bound, for a distributed matrix multiplication job over a network of $18$ nodes.}
   \label{fig:region}
\end{figure}

As an example, in Fig.~\ref{fig:region} we have illustrated the tradeoff between ``computation latency'' and ``communication load'' that is achieved by the unified framework for running a distributed matrix multiplication over $18$ edge nodes (see~\cite{LMA16_unify} Section~III for details). We observe that the achieved tradeoff approximately exhibits an inverse-linearly proportional relationship between the latency and the load. In particular, we can see that the Minimum Bandwidth Codes and the Minimum Latency Codes can be viewed as special instances of the proposed coding framework, by considering two extremes of this tradeoff: minimizing either the communication load or the computation latency individually. Next, we further illustrate how to utilize this tradeoff to minimize the total response time that is the sum of the communication time in the Shuffle phase and the computation latency in the Map phase. For the matrix multiplication problem in Fig.~\ref{fig:region}, we consider real entries each represented using $2$ bytes, a shift-exponential distribution for the Map task execution time, and a wireless network with speed 10 Mbps. Then, the Minimum Bandwidth Codes that wait for all 18 nodes to finish their Map tasks achieve a total response time of 302s,\footnote{The communication load in Fig.~\ref{fig:region} is normalized by the number of the rows of the matrix, which is $10^6$ in this example.} and the Minimum Latency Codes that terminate the Map phase when the fastest 3 nodes (minimum required number) finish their Map tasks achieve a total response time of 263s. Using the unified coding framework, we can wait for the optimal number of the fastest 12 nodes to finish, and achieve the minimum total response time of 186s. Hence, this unified coding approach provides a performance gain of 38.4\% and 29.3 \% over the Minimum Bandwidth Codes and the Minimum Latency Codes respectively.

This unified coding framework, which is essentially a systematic concatenation of the Minimum Bandwidth Codes and the Minimum Latency Codes, takes advantage of both coding techniques in difference stages of the computation. In the Map phase, MDS codes are employed to create coded tasks, which are then assigned to edge nodes in a specific repetitive pattern for local execution. According to the interested computation latency of the Map phase, all running Map tasks are terminated as soon as a certain number of nodes have finished their local computations. Then in the Shuffle phase, coded multicast opportunities specified by Minimum Bandwidth Codes are greedily utilized, until the data demands of all nodes are satisfied. For example, we can consider executing a linear computation consisting of $m=20$ Map tasks using $K=6$ edge nodes, each of which can process $\mu = \frac{1}{2}$ fractions of the tasks.  To be able to end the Map phase when only the fastest $q=4$ nodes finish their local tasks, we can first use a $(\frac{K}{q}m,m)=(30,20)$ MDS code to generate $30$ coded tasks, each of which is then assigned to $\mu q=2$ nodes for execution according to the repetitive assignment pattern specified by the Minimum Bandwidth Codes. For more detailed illustrative examples, we refer the interested readers to Section~IV of~\cite{LMA16_unify}.

The unified coding framework allows us to flexibly select the operation point to minimize the overall job execution time. For example, when the network is slow, we can wait for more nodes to finish their Map computations, creating better multicast opportunities to further slash the amount of data movement. On the other hand, when we have detected that some nodes are running slow or becoming irresponsive, we can shift the load to the network by ending the Map phase as soon as enough coded tasks are executed. 

\section{Conclusions and Future Research Directions}
We demonstrated how coding can be effectively utilized to leverage abundant computing resources at the network edge to significantly reduce the bandwidth consumption and computation latency in Fog computing applications. In particular, we illustrated two recently proposed coding concepts, namely Minimum Bandwidth Codes and Minimum Latency Codes, and discussed their impacts on Fog computing. We also discussed a unified coding framework that includes the above two coding techniques as special cases, and enables a tradeoff between computation latency and communication load to optimize the system performance. 

We envision codes to play a fundamental role in Fog computing by enabling an efficient utilization of  computation, communication, and storage resources at network edge. This area opens up many important and exciting future research directions. Here we list a few:

{\bf Heterogeneous computing nodes:} In distributed Fog networks, different nodes have different processing and storage capacities. The ideas outlined in this paper can be used to develop heuristic solutions for heterogeneous networks. For example, one simple approach is to break the more powerful nodes into multiple smaller virtual nodes that have homogeneous capability, and then apply the proposed coding techniques for the homogeneous setting. However, systematically developing practical task assignment and coding techniques for these systems, that are provably optimum (approximately), is a challenging open problem. 

{\bf Networks with multi-layer and structured topology:} The current code designs for distributed computing\cite{LMA_all,li2016fundamental,LMA16_unify} are  developed for a basic topology, in which the processing nodes are connected through a shared link. While these results demonstrate the significant gain of coding in distributed Fog computing, we need to extend these ideas to more general network topologies. In such networks, nodes can be connected through multiple switches and links in different layers with different capacities. 

{\bf Multi-stage computation tasks:} Another important direction is to consider more general computing frameworks, in which the computation job is represented by a Directed Acyclic Task Graph (DAG). While we can apply the aforementioned code designs for each stage of computation locally, we expect to achieve a higher reduction in bandwidth consumption and response time by globally designing codes for the entire task graph and accounting for interactions between consecutive stages. 

{\bf Coded computing overhead:} The current Fog computing system under consideration lacks appropriate modeling of the coding overhead, which includes the cost for the encoding and decoding processes, the cost for performing multicast communications, and the cost for maintaining desired data redundancy across Fog nodes. To make the study of coding in practical Fog systems more relevant, it is important to carefully formulate a comprehensive model that systematically accounts for these overhead.

{\bf Verifiable distributed computing:} Fog architecture facilitates offloading of computational tasks from relatively weak computational devices (clients) to  more powerful nodes in the edge network. As a result, there is a critical need for ``Verifiable Computing'' methods, in which clients can make sure they receive the correct calculations. This is typically achieved by injecting redundancy in computations by the clients. We expect codes to provide much more efficient methods for leveraging computation redundancy in order to provide verified computing in Fog applications.  

{\bf Exploiting the algebraic structures of computation tasks:} Recall that the Minimum Bandwidth Codes can be applied to any general computation task that can be cast in a MapReduce framework. However, we expect to improve the overall performance, if we exploit the specific algebraic properties of the underlying tasks. For example, if the task has some linearity, we may be able to incorporate it in communication and coding design in order to further reduce the bandwidth consumption and latency. On the contrary, Minimum Latency Codes work only for some particular linear functions (e.g., matrix multiplication). It is of great interest to extend these codes to a broader class of computation tasks. 

{\bf Communication-heavy applications}: Recall that by exploiting Minimum Bandwidth Codes we can envision a Fog system that can handle many distributed Fog nodes with a bounded communication load.  Such a surprising feature would enormously expand the list of applications that can be offered over Fog networks. One research direction is to re-examine some communication-heavy tasks to see if Minimum Bandwidth Codes allow them to be implemented over distributed Fog networks. 

{\bf Plug-and-Play Fog nodes:} We can finally envision a software package (or App) that can be installed and maintained distributedly on each Fog node. This package should allow a Fog computing node to join the system anytime to work with the rest of the nodes or leave the system asynchronously, still the entire network operates near optimum. Designing codes that guarantee integrity of computations despite such network dynamics is a very interesting and important research direction. 

\bibliographystyle{IEEEtran}
\bibliography{ref-abb}

\end{document}